\documentclass[reprint, aps, superscriptaddress, pra]{revtex4-2}
\usepackage[utf8]{inputenc}
\usepackage{graphicx}
\usepackage{amsmath,amssymb}
\usepackage{xcolor}
\usepackage{physics}
\usepackage{siunitx}
\usepackage[version=4]{mhchem}
\usepackage{makecell}
\usepackage{array}
\DeclareSIUnit{\atm}{atm}

\usepackage{hyperref}
\usepackage[capitalize]{cleveref}

\usepackage{braket}
\usepackage{tabularx,booktabs,multirow}

\newcommand{\ntl}{\ce{^{12}C^{12}C^{1}H}}
\newcommand{\dut}{\ce{^{12}C^{12}C^{2}H}}
\newcommand{\dbl}{\ce{^{13}C^{13}C^{1}H}}
\newcommand{\cs}{\ce{^{13}C^{12}C^{1}H}} 
\newcommand{\cw}{\ce{^{12}C^{13}C^{1}H}} 

\AtBeginDocument{
  
}

\begin{document}

\title{Giant Isotope Effect on the Excited-State Lifetime and Emission Efficiency of the Silicon T Centre}

\author{Moein Kazemi}
\thanks{These two authors contributed equally.}
\affiliation{Department of Physics, Simon Fraser University, Burnaby, British Columbia, Canada}
\affiliation{Photonic Inc., Coquitlam, British Columbia, Canada}

\author{Mehdi Keshavarz}
\thanks{These two authors contributed equally.}
\affiliation{Department of Physics, Simon Fraser University, Burnaby, British Columbia, Canada}
\affiliation{Photonic Inc., Coquitlam, British Columbia, Canada}

\author{Mark E. Turiansky}
\affiliation{US Naval Research Laboratory, 4555 Overlook Avenue SW, Washington, DC 20375, USA}

\author{John L. Lyons}
\affiliation{US Naval Research Laboratory, 4555 Overlook Avenue SW, Washington, DC 20375, USA}

\author{Nikolay V. Abrosimov}
\affiliation{Leibniz-Institut für Kristallzüchtung, Berlin 12489, Germany}

\author{Stephanie Simmons}
\affiliation{Department of Physics, Simon Fraser University, Burnaby, British Columbia, Canada}
\affiliation{Photonic Inc., Coquitlam, British Columbia, Canada}

\author{Daniel B. Higginbottom}
\email{daniel\_higginbottom@sfu.ca}
\affiliation{Department of Physics, Simon Fraser University, Burnaby, British Columbia, Canada}
\affiliation{Photonic Inc., Coquitlam, British Columbia, Canada}

\author{Mike L. W. Thewalt}
\affiliation{Department of Physics, Simon Fraser University, Burnaby, British Columbia, Canada}

\date{\today}

\begin{abstract}
Efficient single-photon emitters are desirable for quantum technologies including quantum networks and photonic quantum computers. We investigate the T centre, a telecommunications-band emitter in silicon, and find a strong isotope dependence of its excited-state lifetime. In particular, the lifetime of the deuterium T centre is over five times longer than the common protium variant. Through explicit first-principles calculations, we demonstrate that this dramatic difference is due to a reduction in the carbon-hydrogen local vibrational mode energy, which suppresses non-radiative decay. Our results imply that the deuterium T centre approaches unit quantum efficiency, enabling more efficient single-photon sources, quantum memories, and entanglement generation.
\end{abstract}

\maketitle

\section{Introduction}
\label{sec:introduction}
Solid-state colour centres are among the most promising platforms for quantum information processing and networking~\cite{Weber2010, Aharonovich2016, Awschalom2018, Atature2018, Bradac2019}. They may be single atomic impurities or molecular complexes embedded within a host crystal. Nuclear spins intrinsic to the centre or in external lattice sites can be long-lived quantum memory qubits, and may be entangled with travelling photonic qubits through spin-selective optical transitions \cite{Knaut2024EntanglementNetwork}. Isotopic purification of the host material enhances the optical and spin properties~\cite{Steger2008ReductionConstituents, Balasubramanian2009UltralongDiamond, Saeedi2013_corrected, Christle2015IsolatedTimes., Chartrand2018}
while engineering the isotopic composition of the defect itself tailors the hyperfine structure~\cite{Fuchs2008Excited-stateDiamond, Jacques2009DynamicTemperature, Smeltzer2009RobustDiamond, Felton2009HyperfineDiamond, Steiner2010UniversalDiamond, Pfaff2014}.

The T centre, a carbon-hydrogen complex in silicon, is a promising candidate for practical quantum technologies due to its telecommunications-band emission, paramagnetic ground state, and compatibility with silicon photonics~\cite{Afzal2024, Bergeron:2020_PRX, Higginbottom2022, Deabreu2023waveguide}. The atomic structure~\cite{Safonov1996b}, illustrated as an inset in \cref{fig:pl_spectra}, consists of two inequivalent carbon atoms (\ce{C_S} and \ce{C_W}) and a hydrogen atom (\ce{H}) bonded to \ce{C_W}. In its ground state, the T centre has an unpaired electron localized as a dangling bond on \ce{C_S}. Compared to popular colour centres in wide bandgap semiconductors, and even to other noteworthy emitters in silicon \cite{Lefuacher2023Cavity}, the T centre is relatively inefficient, with an estimated quantum efficiency of 23\% \cite{Johnston_2024_CavityCoupledTcentre}. Non-radiative decay channels present a challenge for T centre quantum technologies by reducing single-photon generation efficiency and optical cyclicity.

Each atomic component of the T centre can be one of several isotopic variants, i.e., hydrogen in its protium (\ce{^1H}) or deuterium (\ce{^2H}) form, and each of the carbon atoms as either \ce{^{12}C} or \ce{^{13}C}.  
The `natural' isotopic form \ce{^{12}C^{12}C^{1}H} is the dominant species in samples with natural isotope distributions. Previous studies observed isotope-dependent shifts in the zero-phonon line (ZPL) and local vibrational mode (LVM) energies, which served as evidence for the proposed defect structure~\cite{Bergeron:2020_PRX, Lightowlers1994d, Safonov1994, Safonov1995, Safonov1996b, Safonov1999a}.

In this work, we report a previously unrecognized isotopic effect: a strong dependence of the excited-state lifetime on the isotopic configuration. In particular, we observe that the lifetime of deuterium T centres is more than five times larger than their protium counterparts. We propose that this significant change arises from differences in the LVM energies among the isotopic variants, analogous to the beneficial kinetic isotope effect that increases the external efficiency of deuterated organic LEDs \cite{Chun2007EnhancementOLED}. Since this difference is orders of magnitude larger than any possible direct isotopic effect on the radiative lifetime, it suggests that the deuterium T centre is strongly radiative---improving the fundamental efficiency of the T centre beyond all prior reports. This is a novel mechanism through which isotopic engineering of the defect itself can directly modify the optical dynamics of solid-state quantum emitters, adding a new dimension to the design of efficient single-photon sources and spin-photon interfaces.   


\section{Isotopic Variants of the T Centre}
In addition to the natural isotopic composition (\ntl), we examine four T centre isotopic variants: `Deuterium' (\dut), `Double \ce{^{13}C}' (\dbl), `Weak  \ce{^{13}C}' (\cw), and `Strong \ce{^{13}C}' (\cs), where the single \ce{^{13}C} variants are named Weak/Strong for their \ce{^{13}C} hyperfine coupling strength to the T centre's electron. These isotopic substitutions embed additional long-lived nuclear spins into the defect, enhancing its potential for quantum information applications. We will show that they also modify the optical properties of the emitter. 

To create these isotopic variants, three bulk samples (A--C) of isotopically enriched $^{28}$Si were prepared. Sample A is an isotopically purified $^{28}$Si crystal previously reported in Ref.~\cite{Bergeron:2020_PRX}. Samples B and C were cut from a FZ $^{28}$Si crystal grown with increasing \ce{^{13}C} concentration along its length. Sample B was further diffused with deuterium gas at \SI{1000}{\celsius} under \SI{1}{\atm} pressure. The new samples were subsequently irradiated with electrons at an energy of \SI{10}{\MeV} to a total dose of \SI{320}{\kilo\gray}, followed by a final thermal annealing step: Sample A was annealed in hydrogen gas from 300--\SI{450}{\celsius} in 30-minute steps, while Samples B and C were annealed in nitrogen gas at \SI{410}{\celsius} for 3 minutes and in open air at \SI{420}{\celsius} for 30 minutes, respectively. For all presented measurements, the samples are loosely mounted in strain-free reflective pockets and immersed in liquid helium at either 1.4 or \SI{4.2}{\kelvin}. 

\begin{figure}[t]
  \centering
  \includegraphics[width=\linewidth]{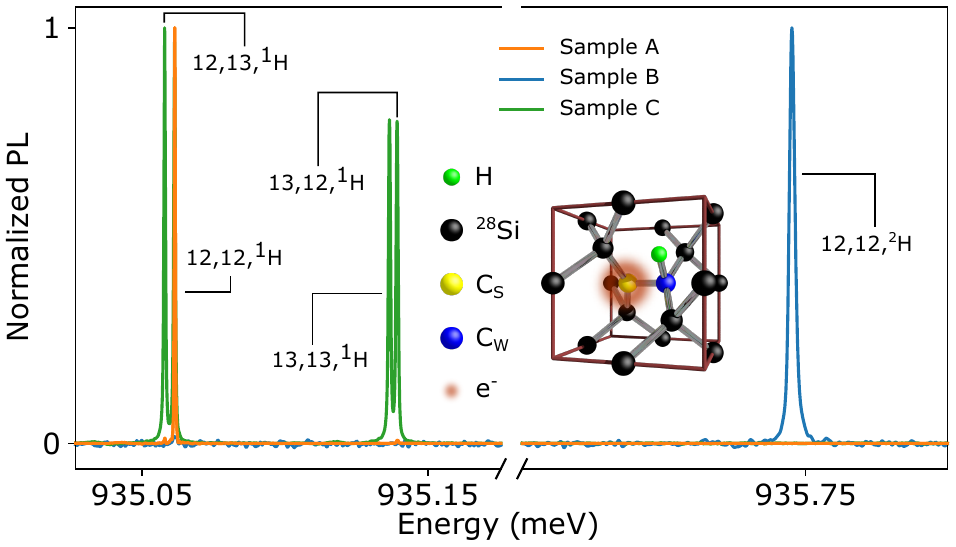}
  \caption{Photoluminescence spectra of T centre samples with varying isotopic composition. Each peak is labelled with the attributed structure, following a \ce{C_S},\ce{C_W},\ce{H} labelling scheme. Sample A (orange) is the natural isotope distribution. Sample B (blue) is deuterated. Sample C (green) is grown with an elevated concentration of \ce{^{13}C}. (Inset) Atomic structure of the T centre showing the inequivalent carbon sites \ce{C_S} (dangling bond) and \ce{C_W} (bonded to hydrogen).}
  \label{fig:pl_spectra}
\end{figure}

We excite the samples using an above-bandgap laser to measure the non-resonant photoluminescence (PL). \Cref{fig:pl_spectra} shows the above-band PL spectra of Samples~A, B, and C, taken at a temperature of \SI{1.4}{\kelvin}. Sample~A, composed primarily of \ce{^{12}C} and not deliberately deuterated, exhibits the natural T centre emission. In Sample~B, which is deuterated and has natural (1.1\%) \ce{^{13}C} fraction, the dominant variant is the deuterium T centre. A small contribution from the natural T centre remains visible in Sample~B due to residual protium introduced during growth or processing. Sample~C, grown with nearly equal \ce{^{12}C} and \ce{^{13}C}, features all four carbon isotopic variants in approximately equal proportions. The ZPL shifts of the isotope variants compared to the natural T centre are listed in \cref{tab:lifetime} and consistent with previously reported measurements \cite{Safonov1994,Bergeron:2020_PRX}.

\section{Lifetime Measurements}

\begin{figure}[t]
  \centering
  \includegraphics[width=\linewidth]{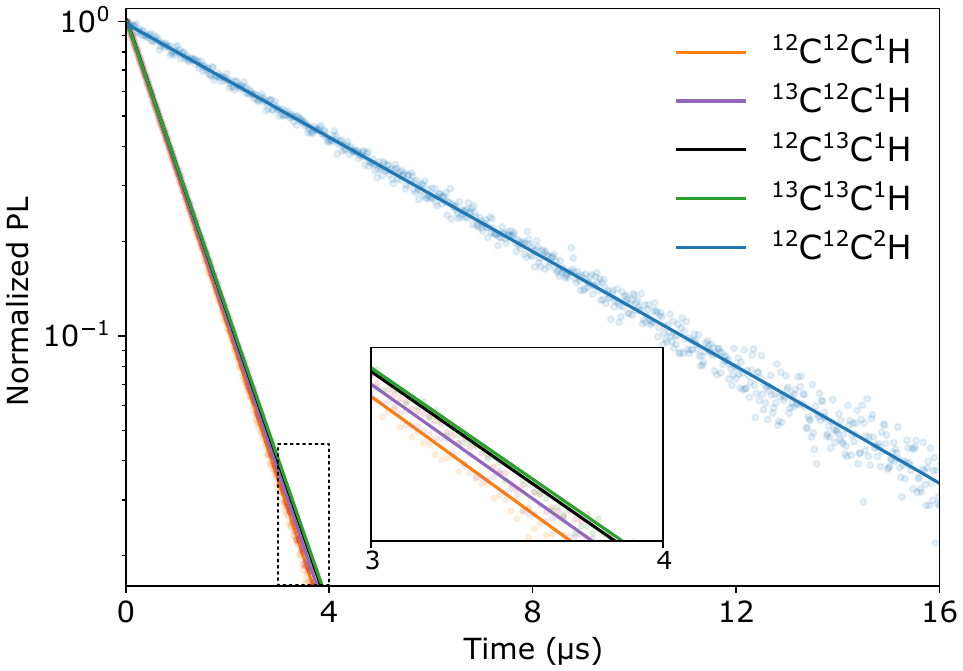}
  \caption{Luminescence transient decay of T centre isotope variants under pulsed resonant excitation. The inset shows an enlarged view of the region enclosed by the dashed box.}
  \label{fig:resonant_lifetime}
\end{figure}

We measure the excited-state lifetime of each isotopic variant in samples~B (deuterated) and C (\ce{^{13}C}-enriched) at \SI{4.2}{\kelvin} under both non-resonant and resonant excitation. Resonant excitation is performed using short pulses of $\sim \SI{1326}{\nm}$ laser light tuned exactly to the ZPL frequency, determined by photoluminescence excitation (PLE) spectroscopy. Resonant lifetimes are obtained by pulsing the resonant laser and recording the transient luminescence of the phonon and LVM sideband. \Cref{fig:resonant_lifetime} shows the excited-state decay of T centre isotope variants, with all measured lifetimes given in \cref{tab:lifetime}.

\begin{table}[h]
\centering
\caption{ZPL isotope shifts and excited-state lifetimes of T centre isotopic variants under resonant excitation.}
\begin{ruledtabular}
\begin{tabularx}{\columnwidth}{l l c c}
Variant & Structure & ZPL shift (\si{\micro\eV}) & Lifetime (\si{\micro\second}) \\
\hline 
Natural          & \ntl  &  ---     & \num{0.885 \pm 0.004} \\
Strong \ce{^{13}C} & \cs  & +78.04  & \num{0.904 \pm 0.001} \\
Weak \ce{^{13}C}   & \cw  & $-3.47$ & \num{0.921 \pm 0.001} \\
Double \ce{^{13}C} & \dbl & +75.28  & \num{0.929 \pm 0.001} \\
Deuterium         & \dut & +745    & \num{4.807 \pm 0.018} \\
\end{tabularx}
\end{ruledtabular}
\label{tab:lifetime}
\end{table}

Deuterium T's significantly ($5.4 \times$) larger lifetime stands out. Based on the T centre radiative efficiency estimates to date \cite{Higginbottom2022,Johnston_2024_CavityCoupledTcentre}, a change to zero-phonon or phonon-assisted radiative decay alone cannot account for a lifetime modification of this magnitude. We also measure the excited-state lifetime of deuterium T centres in Sample B under above-bandgap excitation using a \SI{980}{\nano\metre} diode laser. The lifetimes of deuterium T obtained under resonant and above-bandgap excitation agree within 1\%. For the natural T centre in the same sample, the above-bandgap lifetime, is approximately \SI{0.967 \pm 0.025}{\micro\second}, consistent with previous bulk and waveguide-integrated centres~\cite{Bergeron:2020_PRX,Deabreu2023waveguide}. Across samples, variations of up to \SI{100}{\nano\second} in the natural T centre lifetime are observed under above-bandgap excitation, likely due to sample-dependent free-exciton capture times. In contrast, resonant excitation yields a consistent lifetime of \SI{0.885(4)}{\micro\second} for the natural T centre. 

In addition, we measure the excited-state lifetimes of deuterium M and I centres~\cite{Safonov1997a, Gower1997a, Filippatos2025Re-examinationComputing, Filippatos2025FirstPrinciplesFunctionals, Kuganathan2025Re-examiningSilicon}, T-like hydrogen-containing luminescence centres in silicon. In all cases, a consistent and substantial increase in lifetime is observed for the deuterium variants compared to their protium counterparts. The details of these measurements will follow in a forthcoming publication.

\section{Theory}
The excited-state lifetime ($\tau$) is determined by both radiative and nonradiative decay channels:
\begin{equation}
    \label{eq:lifetime}
    \frac{1}{\tau} = \Gamma_{\rm NR} + \Gamma_{\rm R} \;,
\end{equation}

where $\Gamma_{\rm NR}$ is the nonradiative decay rate and $\Gamma_{\rm R}$ is the radiative decay rate.
Within the Franck-Condon approximation, the dipole matrix element that enters the evaluation of the radiative rate is purely electronic~\cite{Stoneham1975} and, to lowest order, the radiative rate can be assumed to be independent of the isotopic composition. Indeed, the measured lifetime of deuterium T is within 4\% of longest radiative lifetime (\SI{5}{\us}) given in Ref.~\cite{Dhaliah2022}, suggesting that the non-radiative decay rate is substantially reduced. We therefore attribute the observed lifetime isotope dependence of the T centre to a modification of the nonradiative decay channels.

Among possible nonradiative channels~\cite{Stoneham1975}, the process enabled by multiphonon emission~\cite{Alkauskas2014} often dominates. Of the plausible alternatives, earlier measurements were unable to identify Auger-Meitner recombination from the T centre \cite{Bergeron:2020_PRX}.
Since the multiphonon process depends exponentially on system parameters, including the phonon energy~\cite{Englman1970TheMolecules,Turiansky2024}, it may explain the observed isotopic lifetime variation.
We test this hypothesis with first-principles calculations of the nonradiative transition rate using parameters from hybrid density functional theory (see \cref{sec:computational_details} for details of the calculation).
Alkauskas \textit{et al.} proposed a single-mode approximation to evaluate the rate~\cite{Alkauskas2014}, in which the chosen mode connects the ground and excited state geometries and is known as the ``accepting mode''.
Assuming $T = 0$, which is valid for the low temperatures of this study, the nonradiative decay rate is given by
\begin{multline}
    \label{eq:nr}
    \Gamma_{\rm NR} = \frac{2\pi}{\hbar} \lvert W_{eg} \rvert^2 \times \\ \sum_n {\lvert \braket{\chi_{e0} \lvert \hat{Q} - Q_0 \rvert \chi_{gn}} \rvert}^2 \, \delta (E_{\rm ZPL} - n\hbar\Omega_g) \;,
\end{multline}
where $W_{eg}$ is the electron-phonon coupling matrix element, $\chi_{e/g,n}$ are the harmonic oscillator wavefunctions for the ground ($g$) or excited ($e$) state with vibrational frequency $\Omega_{e/g}$, and $\hat{Q}$ is the phonon position operator evaluated with respect to the equilibrium geometry $Q_0$.
We evaluate the rate using the Nonrad code~\cite{Turiansky2021}.

\begin{figure}[b]
    \centering
    \includegraphics[width=\columnwidth,height=0.5\textheight,keepaspectratio]{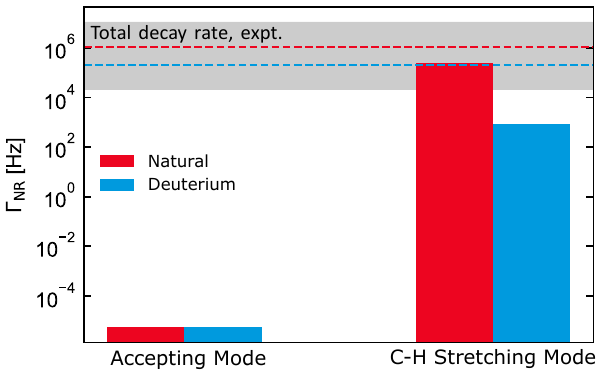}
    \caption{\label{fig:nr}
        The nonradiative decay rate $\Gamma_{\rm NR}$ of the natural (red) and deuterium (blue) T centre evaluated for the accepting and C-H stretching modes.
        The dashed lines correspond to the experimental lifetimes of \SI{0.884}{\micro\second} (red) and \SI{4.807}{\micro\second} (blue), and the gray, shaded region highlights the range within one order of magnitude of the experimental values.
    }
\end{figure}

\Cref{fig:nr} shows the calculated nonradiative rates for natural and deuterium T centres.
The accepting mode approximation fails dramatically: the calculated rates fall 10 orders of magnitude below experiments and show no isotopic variation.
For the T centre, the accepting mode corresponds to breathing of the four Si atoms bonded to the two C atoms of the defect, yielding a \SI{33.0}{\meV} phonon energy which is within the range of bulk Si phonon modes.
The absence of H motion explains the hydrogen isotope independence.
Despite its wide utilization as the basis of first-principles calculations, the accepting mode approximation lacks formal justification~\cite{Alkauskas2014}.

In contrast to systems where the accepting mode approximation has been successfully applied, the T centre possesses LVMs with larger energy than the bulk Si phonon modes~\cite{Safonov1996b}.
The protium C-H stretching mode, in particular, has a predicted phonon energy of $\approx$\SI{361}{\meV}, and replacing hydrogen with the heavier deuterium reduces the energy to $\approx$\SI{265}{\meV}~\cite{Safonov1996b}.
Both energies are significantly larger than the accepting mode energy.
Given the exponential dependence of the nonradiative rate on phonon energy~\cite{Englman1970TheMolecules,Turiansky2024}, we introduce the \textit{ansatz} that the C-H stretching mode plays an essential role in enabling the nonradiative decay of the T centre.

We re-evaluate the nonradiative rate using the C-H stretching mode in lieu of the accepting mode and obtain the results shown in \Cref{fig:nr}.
The calculated nonradiative rate now agrees well with the experimental lifetime, to within the computational uncertainty.
In addition, the nonradiative rate for the natural T centre is $\approx$285$\times$ larger than that of the deuterium T centre, rationalizing the significant dependence of the experimentally measured excited state lifetime on the hydrogen isotope.

\section{Discussion}

The giant isotope effect on the lifetime of the silicon T centre that we measure is analogous to the observed isotope effects of deuterated organic molecules \cite{Ma2023ImprovingLuminescence} and organic light emitting diodes \cite{Chun2007EnhancementOLED}. In both deuterated emitters and hosts, lifetime and efficiency enhancements are attributed to a beneficial kinetic isotope effect which suppresses motional dissipation, yielding greater stability and operational lifetime \cite{Yuan2025HighEfficiency}. Similarly, the significant isotope effect of hot-carrier-induced dissociation of Si-H bonds at Si/SiO$_2$ interfaces has led device manufacturers to use deuterium in place of hydrogen in metal-oxide-semiconductor transistors~\cite{Lyding1996}. To our knowledge, this is the first observation of a related isotope-lifetime effect in a semiconductor colour centre.

The measured isotope-lifetime dependence of the T centre has important implications for both quantum device engineering and semiconductor defect modelling. Theoretical estimates of its radiative lifetime vary from \SI{1.3}{\micro\second} to \SI{5}{\micro\second} \cite{Dhaliah2022,Filippatos2025FirstPrinciplesFunctionals, MarkCN}.
From \cref{eq:lifetime}, the assumed isotopic insensitivity of the radiative rate, and the experimentally measured lifetimes, we estimate a radiative lifetime of $\approx$\SI{4.9}{\micro\second}.
This implies a quantum efficiency of $\eta_\mathrm{H}\approx\SI{18.1}{\percent}$ for the natural T centre and $\eta_\mathrm{D}\approx\SI{98.4}{\percent}$ for the deuterium variant. Given the measured PSB fraction $\eta_\mathrm{DW} = 0.23$ \cite{Bergeron:2020_PRX}, this corresponds to an improvement in total zero-phonon radiative emission from 4\% to 23\%. 
Ref.~\onlinecite{Johnston_2024_CavityCoupledTcentre} reported a T centre's quantum efficiency $\eta$ exceeding \SI{23.4}{\percent}.
Even if the deuterium T centre was perfectly radiative, the implied efficiency of the natural variant would not exceed $\approx$\SI{18.4}{\percent}. Further measurements of radiative efficiency and isotope-dipole dependence may resolve this discrepancy.

These efficiency estimates are subject to uncertainty in the isotopic dependence of the nonradiative rate extracted from calculations.
Although our simple \textit{ansatz} dramatically outperforms the typical accepting mode approach, a multimode treatment (the subject of future work) is expected to further improve accuracy. Furthermore, the calculation is sensitive to the C-H stretch mode energy, which has not previously been measured for either the protium or deuterium T centres. As described in \cref{sec:LVM_measurements}, we observe a PSB feature tentatively attributed to the natural T centre C-H stretch mode. However, we were not able to observe the corresponding deuterium mode in this study due to a weaker T signal in that sample.

These results present a dramatic increase in the performance of silicon T centre devices for quantum networking and computing over earlier proposals \cite{Yan2021,Higginbottom2023memory,Taherizadegan2025ExploringFeasibility,Khalifa2025RobustMicrowave}. Quantum efficiency is one of the critical metrics for solid-state emitters. Compared to their protium counterparts, deuterium T centre single-photon sources can be expected to offer superior efficiency, with quadratic improvement to the bipartite remote spin entanglement rate \cite{Afzal2024}. Similar efficiency boosts can be expected for deuterium T ensemble optical quantum memories and microwave to optical transducers \cite{Higginbottom2023memory,Khalifa2025RobustMicrowave}.

In nanocavity devices with high Purcell factor $P$, where enhanced radiative emission dominates all other decay processes (irrespective of isotopic variant), the efficiency difference will be marginal. However, even in this regime, suppressing nonradiative decay may increase electron spin cyclicity dramatically. Models of the T centre excited state predict highly cyclic radiative decay \cite{Clear2024optical}, whereas nonradiative decay can be treated as comparatively (if not fully) spin-mixing. From \cref{eq:lifetime}, considering only non-radiative spin mixing and assuming complete mixing, the electron cyclicity is 
\begin{equation}
C_\mathrm{NR} = \frac{2}{1-\eta_\mathrm{rad}(P)} = 2 \left( \frac{1+\eta_0 (P - 1) }{1-\eta_0} \right)\,,
\end{equation}
where $\eta_0$ is the intrinsic (zero-Purcell) radiative efficiency. In the high-$P$ limit, the cyclicity improvement between protium and deuterium T centre devices is a factor of $278$, assuming the radiative efficiencies above. Thus, replacing protium with deuterium makes quantum non-demolition readout of the T centre electron spin significantly more tolerant to optical path losses, bringing high-fidelity single-shot readout within reach. Unless protium's singlet nuclear spin or larger gyromagnetic ratio is required, the deuterium T centre is broadly advantageous.

\section{Conclusion}

We have reported previously unobserved excited state lifetime variations for the isotopic variants of the silicon T centre. In particular, the deuterium T centre exhibits a lifetime over five times longer than the common protium version. We attribute this to suppressed nonradiative decay rate due to the decreased energy of the C-H stretching mode when protium is replaced with deuterium. Although the standard accepting mode model of multiphonon decay fails to capture this isotope dependence, we find that a simple model considering only the C-H stretching mode shows not only strong isotope dependence but also accounts for the observed excited state lifetime to reasonable computational uncertainty. We estimate that the deuterium T centre's emission efficiency is $\approx 98.4\%$, among the highest for native silicon emitters. This variant is expected to outperform the protium variant for quantum information applications, including single-photon sources, quantum memories, repeaters, transducers, and spin-photon quantum computing architectures.

\section{Acknowledgments}
    
This work was supported by the Natural Sciences and Engineering Research Council of Canada (NSERC) through Discovery Grants held by D.B.H, S.S., and M.L.W.T., the Canadian Department of National Defence (DND) through an IDEAS Micronet, the New Frontiers in Research Fund (NFRF), the Canada Research Chairs program (CRC), the Canada Foundation for Innovation (CFI), the B.C. Knowledge Development Fund (BCKDF), and the Canadian Institute for Advanced Research (CIFAR) Quantum Information Science program. 

M. Kazemi and M. Keshavarz are supported by the NSERC CREATE program QSciTech.
    
    M.\,E.\,T. and J.\,L.\,L. were supported by the Office of Naval Research through the Naval Research Laboratory's Basic Research Program.

    Sample A was prepared from the Avo28 crystal produced by the International Avogadro Coordination (IAC) Project (2004--2011) in cooperation among the BIPM, the INRIM (Italy), the IRMM (EU), the NMIA (Australia), the NMIJ (Japan), the NPL (UK), and the PTB (Germany).

    We acknowledge helpful discussions with Alexander Patscheider and Evan MacQuarrie of Photonic Inc., who privately shared corroborating data.

\appendix

\section{Experimental methods}
\label{sec:exp_methods}

PL and LVM spectra are acquired using a Bruker IFS 125 HR Fourier transform infrared (FTIR) spectrometer equipped with a \ce{CaF_{2}} beam splitter and either a liquid-nitrogen-cooled \ce{Ge} diode detector (for ZPL PL) or InSb detector (for the LVM PL in \cref{fig:4}). The liquid-nitrogen cooled InSb detector includes a cooled Schott RG850 and Spectrogon \SI{2600}{\nano\metre} short-pass filter to block black-body background radiation. The spectral resolution of the ZPL PL spectrum is \SI{0.25}{\micro\eV} for Samples~A and~C and \SI{0.62}{\micro\eV} for Sample~B. The LVM PL spectrum shown in \Cref{fig:4} is measured at a resolution of \SI{0.124}{\milli\eV}. 

Resonant excitation for both lifetime and LVM measurements is provided by a Toptica continuously tunable diode laser, amplified with a Toptica BoosTA Pro, and frequency-stabilized to a Bristol 871 wavemeter. The excitation beam passes through a \SI{1325}{\nano\metre}\SI{\pm 2.5}{\nano\metre} band-pass filter. For LVM studies, a \SI{1500}{\nano\metre} long-pass filter is used to reject residual excitation light on the collection side. For lifetime measurements, the laser is pulsed using a fibre-coupled AeroDiode AOM, and the luminescence is filtered through three \SI{1319}{\nano\metre} filters, followed by a \SI{1425}{\nano\metre} notch filter, whose purpose is to block the zone-centre optical phonon Raman line. Luminescence is detected with an ID Quantique ID230 InGaAs single-photon detector operating at 10\% efficiency with a \SI{15}{\micro\second} dead time. For above-bandgap lifetime measurements, a pulsed \SI{980}{\nano\metre} diode laser is used.

\section{LVM measurements}
\label{sec:LVM_measurements}
\begin{figure}[t h!]
    \centering
    \includegraphics[width=\columnwidth,height=0.5\textheight,keepaspectratio]{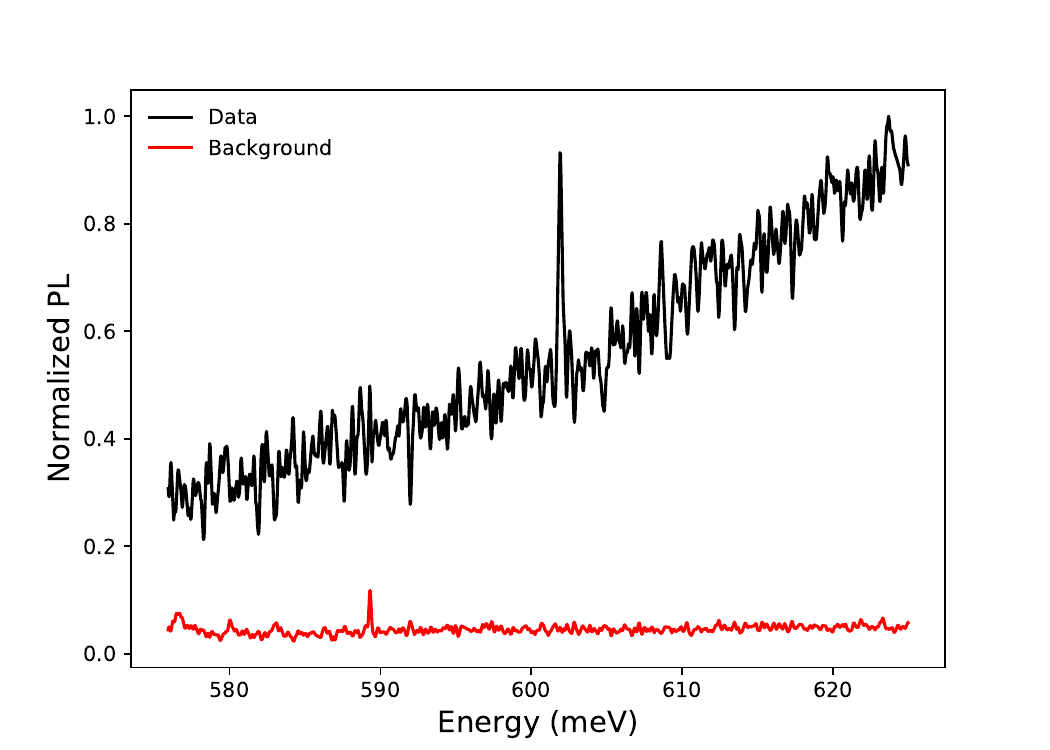}
    \caption{\label{fig:4}
    Resonant photoluminescence spectrum of the natural T centre (Sample A), showing the region near the expected C–H stretch LVM energy. The black curve shows the background-subtracted and system-response–normalized data, while the red curve shows the background signal.
    }
\end{figure}

The C–H stretch LVM of the T centre has not previously been experimentally observed; the carbon-protium (carbon-deuterium) stretch mode has been predicted by Safonov \textit{et al.}~\cite{Safonov1996b} to have energy $\approx$ \SI{361}{meV} (\SI{265}{meV}). We perform resonant photoluminescence (PL) measurements on Sample~A and identify a candidate C-P LVM feature. A background spectrum is acquired with the laser detuned by \SI{8.27}{\micro\eV} and subtracted from the resonant spectrum. After background subtraction and normalization to the system response, a weak spectral feature appears at \SI{601.9}{\milli\electronvolt}, as shown in \cref{fig:4}, which we tentatively attribute to the C–P stretch LVM, corresponding to an LVM energy of \SI{333.1}{\milli\electronvolt}. However, the corresponding C–D stretch mode for the deuterated T centre is not observed due to the weaker T signal in this sample. A detailed investigation of the LVM spectra across all isotopic variants of the T centre will be presented in a separate publication.

\section{Computational details}
\label{sec:computational_details}

We perform density functional theory calculations with VASP code~\cite{Kresse1996a,Kresse1996b} version 6.4.3.
Core electrons are frozen within the projector augmented-wave formalism~\cite{Blochl1994}, and the valence electrons are represented in a plane-wave basis truncated at an energy of 400~eV.
To enable quantitative predictions, we utilize the hybrid functional of Heyd, Scuseria, and Ernzerhof~\cite{Heyd2003,Heyd2006} with default mixing (25\%) and screening (0.2~{\AA}$^{-1}$), commonly referred to as HSE06.
These choices are consistent with previous studies of the T center~\cite{Dhaliah2022, Turiansky2025}.
Utilizing the standard approach for simulating defects in periodic boundary conditions~\cite{Freysoldt2014}, the T center is investigated in a 512-atom supercell, which is a $4\times4\times4$ multiple of the conventional cubic unit cell, and the Brillouin zone of the supercell is sampled at the $\Gamma$ point.
Atomic coordinates are relaxed until the forces are below 5~meV/{\AA}.
The lattice parameters are held fixed at the calculated bulk value.
We explicitly take spin polarization into account.

Previous work~\cite{Turiansky2025} has demonstrated that the geometry of the bound-exciton excited state is well described by the geometry of the negative charge state, based on the suggestion of Ref.~\onlinecite{Silkinis2025}.
We use the negative charge state to derive the mass-weighted atomic relaxation $\Delta Q$ and phonon frequencies $\Omega_{g/e}$ needed for the evaluation of \cref{eq:nr}.
The electron-phonon coupling matrix element $W_{eg}$ is evaluated in the ground state, and we average over the squared matrix elements of the three degenerate valence-band maxima.
This evaluation of the matrix element takes advantage of the fact that the density of the lowest hydrogenic effective-mass state at the origin is approximately equal to the inverse of the 512-atom supercell volume.
The delta function in \cref{eq:nr} is replaced with a Gaussian with a broadening of $\hbar\Omega_e / 2$.
For the energy difference between the ground and excited state, we utilize the experimental value of 935~meV.
The remaining parameters used in evaluating the nonradiative transition rate are given in \cref{tab:comp_params}.

\begin{table}[h!]
    \centering
    \caption{\label{tab:comp_params}
        The mass-weighted atomic relaxation $\Delta Q$, phonon frequencies $\Omega_{g/e}$ in the ground $g$ and excited $e$ states, and electron-phonon coupling matrix element $W_{eg}$ for the accepting mode and C-H stretching mode.
        Values are given assuming natural isotopic abundances.
        The values from replacing hydrogen with deuterium are given in parentheses.
    }
    \begin{ruledtabular}
        \begin{tabularx}{\columnwidth}{ccc}
            & \multicolumn{2}{c}{Vibrational Mode} \\
            \cline{2-3}
            Parameter & Accepting & C-H Stretch \\
            \midrule
            $\Delta Q$ [amu$^{1/2}$~{\AA}]& 0.734 (0.734) & 0.001 (0.002) \\
            $\hbar\Omega_g$ [meV] & 33.0 (33.0) & 359 (263) \\
            $\hbar\Omega_e$ [meV] & 33.0 (33.0) & 358 (262) \\
            $W_{eg}$ [meV/(amu$^{1/2}$~{\AA})] & 9.23 (9.23) & 0.58 (0.70) \\
        \end{tabularx}
    \end{ruledtabular}
\end{table}

\bibliographystyle{apsrev4-2}
\bibliography{references,extra_references}

\end{document}